\documentclass[%
 reprint,
showpacs,preprintnumbers,
 amsmath,amssymb,
 aps,
prb,
]{revtex4-1}

\usepackage{graphicx}
\usepackage{dcolumn}
\usepackage{bm}
\usepackage{color}
\usepackage{ulem}

\begin{document}

\preprint{APS/123-QED}

\title{Toroidal order in metals without local inversion symmetry 
}


\author{Satoru Hayami,$^1$ Hiroaki Kusunose,$^2$ and Yukitoshi Motome$^1$}
\affiliation{%
 $^1$Department of Applied Physics, University of Tokyo, Tokyo 113-8656, Japan \\
 $^2$Department of Physics, Ehime University, Matsuyama 790-8577, Japan
}%

\begin{abstract}
Toroidal order, given by a composite of electric and magnetic orders, 
manifests itself not only in 
the peculiar magnetism but also in anomalous transport and magnetoelectric effect. 
We report our theoretical 
results on the influence and stability of a toroidal order 
in metals on the basis of a microscopic model.
We consider an effective single-band Hubbard-type model 
with a site-dependent antisymmetric spin-orbit coupling, 
which is derived from a four-band tight-binding model including 
the atomic spin-orbit coupling, off-site hybridizations between orbitals with different parities, 
and {\it odd-parity} crystalline electric field. 
For this single-band model on a layered honeycomb lattice, 
we investigate the electronic structure, magnetotransport, and magnetoelectric effect in the toroidal ordered state with a vortex-like magnetic structure. 
The ferroic order of the microscopic toroidal moments acts 
as an effective gauge field for electrons, which 
modulates the electronic band structure with a shift of the band bottom in the momentum space.
In addition, the site-dependent antisymmetric spin-orbit coupling gives rise to highly anisotropic Hall responses. 
The most salient feature is two different types of magnetoelectric responses: 
one is a magnetic order with net toroidal magnetization induced by 
an electric current perpendicular to the planes, and the other
is a uniform transverse magnetization induced by an electric current within the planes. 
We examine the ground state of the effective model by the mean-field approximation, and 
show that the toroidal order is 
stabilized by strong electron correlations at low electron density. 
We also 
discuss the temperature dependence of the magnetoelectric effects 
associated with spontaneous toroidal ordering. 
Implications to experiments are also presented. 
\end{abstract}
\maketitle


\section{Introduction}

Magnetoelectric effect, which is a consequence of the 
interplay between electric and magnetic properties of electrons first proposed by Curie~\cite{curie1894symetrie}, 
has long been studied extensively 
in condensed matter physics~\cite{Fiebig0022-3727-38-8-R01, Spaldin2005renaissance, 2006multiferroic}. 
It is an intriguing phenomenon where the magnetization is induced by 
an electric field and the electric polarization is induced by a magnetic field. 
Recently, it has attracted renewed interest since the discovery of multiferroic materials showing 
large magnetoelectric responses~\cite{kimura2003magnetic,wang2003epitaxial,hur2004electric}.  
In these magnetic insulators, 
both spatial-inversion and time-reversal symmetries are broken by a spontaneous magnetic order,
yielding simultaneously a uniform electric polarization and a magnetization~\cite{Katsura:PhysRevLett.95.057205,Mostovoy:PhysRevLett.96.067601,SergienkoPhysRevB.73.094434,ArimaJPSJ.80.052001}. 
Such multiferroic materials have been intensively studied not only from the viewpoint of 
fundamental physics but also for potential applications to multifunctional devices~\cite{scott2007data,ramesh2007multiferroics,chu2008electric,KhomskiiPhysics.2.20,Pyatakov1063-7869-55-6-R02}. 

A toroidal order 
is one of the states of matter showing such cross-correlations between 
electric and magnetic responses. 
A toroidal moment, which is represented by a vector product of electric and magnetic moments, 
was originally introduced as an anapole moment in the context of parity violation by weak interactions~\cite{zel1958relation}. 
Recently, a toroidal order, a periodic array of toroidal moments in crystals, 
has gained interest because it leads to exotic phenomena, 
such as a diamagnetic anomaly and nonreciprocal directional dichroism~\cite{gorbatsevich1994toroidal,popov1998magnetoelectric,schmid2001ferrotoroidics,EdererPhysRevB.76.214404,Spaldin:0953-8984-20-43-434203,kopaev2009toroidal,MiyaharaJPSJ.81.023712}, in addition to ordinary magnetoelectric effects.

There are several multiferroic materials which exhibit toroidal orders. 
For instance, Cr$_2$O$_3$ shows a toroidal order in the spin-flop phase 
under strong magnetic field~\cite{FolenPhysRevLett.6.607,popov1999magnetic}. 
In the magnetic piezoelectric material GaFeO$_3$, 
a ferroic toroidal order was detected by resonant magnetoelectric X-ray scattering~\cite{arima2005resonant}. 
LiCoPO$_4$ exhibits a large linear magnetoelectric effect, and the coexistence of ferrotoroidic and antiferromagnetic domains was observed by optical second harmonic generation~\cite{van2007observation}. 
Ba$_2$CoGe$_2$O$_7$ gives rise to a spontaneous toroidal order due to single-ion effects~\cite{ToledanoPhysRevB.84.094421}.
These findings are thus far restricted to insulators.

Toroidal orders can exist also in metallic systems despite the absence of a macroscopic polarization. 
Their influences on electronic and magnetoelectric properties can be more interesting than in insulators owing to the conducting nature. 
For instance, recently, an antiferromagnetic metal on a zigzag lattice, which accommodates a toroidal order, was shown to exhibit an interesting magnetoelectric response~\cite{Yanase:JPSJ.83.014703}. 
Nonetheless, toroidal orders in metals have not been studied intensively, in particular, from the microscopic point of view. 
For further stimulating experiments on toroidal ordered systems, 
it is desired to systematically study how toroidal orders affect 
the electronic structure, transport properties, and magnetoelectric effects. 
It is also important to examine the stability of 
toroidal orders and to clarify the finite-temperature behavior associated with spontaneous toroidal ordering. 

In the present study, 
we investigate a microscopic model in order to clarify the effect of toroidal ordering in metals. 
In particular, we examine the effect of spontaneous ferroic ordering of 
toroidal moments with a focus on the lattice structures on which the spatial-inversion symmetry is preserved globally but broken intrinsically at each magnetic site. 
We consider a low-energy effective single-band model for a minimal four-band tight-binding model. 
The site-dependent antisymmetric spin-orbit coupling in the effective model is derived from the atomic spin-orbit coupling, off-site hybridization between orbitals with different parities, 
and odd-parity crystalline electric field together with electron-electron interactions. 
As a typical example, we study the effect of a toroidal order triggered by a vortex-like magnetic order 
in the effective model on a layered honeycomb lattice. 
The ferroic ordering of 
microscopic toroidal moments acts as 
an effective gauge field, which leads to modulations of the band dispersions with a shift of the band bottom from the $\Gamma$ point. 
We find that the toroidal magnetic order plays an important role in 
the anisotropic magnetotransport and the magnetoelectric effects depending on the direction of the applied electric current. 
In particular, we elucidate that an out-of-plane electric current induces a vortex-like magnetic order, 
while an in-plane current yields a transverse uniform magnetization in the plane by canting the underlying vortex-like magnetic order. 
We discuss such effects from a symmetry point of view. 
We also provide some implications to experiments for toroidal metals. 
Although the qualitatively similar results were obtained for a related model in the previous study~\cite{Yanase:JPSJ.83.014703}, 
the present work provides a comprehensive analysis focusing on the toroidal ordering. 
Moreover, investigating the model by a mean-field approximation, we find that such a ferroic toroidal order appears at low temperatures 
in a wide range of strongly-correlated regions for low electron density. 
We also show that temperature dependence of the anomalous magnetoelectric responses. 
We find that the longitudinal toroidal response to the current exhibits a broad peak around the critical temperature with a kink at the transition, while the transverse uniform magnetization induced by the current shows order-parameter like behavior. 

The organization of this paper is as follows. 
In Sec.~\ref{Sec:Toroidal moment}, we give a brief review on the microscopic definition and the symmetry analysis of toroidal moments. 
In Sec.~\ref{Sec:Toroidal ordering in metals}, we investigate the influence of toroidal ordering in metals. 
After introducing the lattice structures with local inversion symmetry breaking in Sec.~\ref{Sec:Lattices with local inversion symmetry breaking}, we present a low-energy effective Hamiltonian in Sec.~\ref{Sec:Model}. 
We show how the toroidal order affects the electronic structure, anisotropic magnetotransport, and magnetoelectric effects in Secs.~\ref{Sec:Electronic structure}, \ref{Sec:Magnetotransport}, and \ref{Sec:Magnetoelectric effect}, respectively. 
In Sec.~\ref{Sec:Mean field calculations}, we examine the stability of the toroidal magnetic order 
at the level of mean-field approximation. 
The results for the ground state and the finite-temperature properties are presented in Secs.~\ref{Sec:Ground state phase diagram} and \ref{Sec:Finite temperature property}, respectively. 
Section~\ref{Sec:Summary} is devoted to a summary of the present paper.

\section{Toroidal moment}
\label{Sec:Toroidal moment}
\begin{figure}[t!]
\begin{center}
\includegraphics[width=0.6 \hsize]{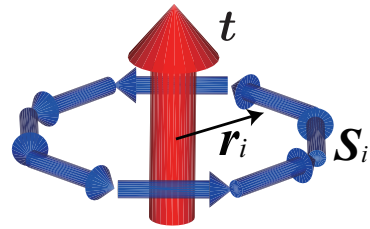} 
\caption{
\label{Fig:Toroidal_structure}
(Color online) 
Schematic picture of the toroidal moment, which is defined by a vector product of the position from inversion center and the magnetic moment.
See Eq.~(\ref{Eq:toroidal_spin}).
}
\end{center}
\end{figure}

In this section, we introduce the toroidal moment from both microscopic and macroscopic points of view. 
Although these arguments were already given in the literatures, e.g., in Ref.~\onlinecite{Spaldin:0953-8984-20-43-434203}, we give a brief summary for making the present paper self-contained and for understanding the microscopic results in the 
following sections. 

First, we discuss a microscopic origin of the toroidal moment. 
In general, magnetic and toroidal multipoles appear in the multipole expansion of an electromagnetic vector potential, 
\begin{align}
\bm{A} &=
\sum_{lm} \sqrt{\frac{4\pi(l+1)}{2l+1}} \frac{1}{r^{l+1}} \nonumber \\ 
\times & \left( \frac{\rm i}{\sqrt{l}}M_{lm} \bm{Y}^{l}_{lm} - \frac{\sqrt{2l+1}}{r} T_{lm} \bm{Y}^{l+1}_{lm} \right), 
\label{eq:A_expansion} 
\end{align}
where $\bm{Y}_{lm}^{l'}$ is 
vector spherical harmonics~\cite{Jackson3rd1999,Blatt1991,Kusunose:JPSJ.77.064710}; 
$l$ and $m$ are the azimuthal and magnetic quantum numbers, respectively, and $l'=l,l\pm1$. 
$\bm{Y}_{lm}^{l'}$ has parity $(-1)^{l'}$ under spatial inversion.
Since $\bm{A}$ is a polar vector with time-reversal odd, 
$M_{lm}$ in the first term of Eq.~(\ref{eq:A_expansion}) represents the magnetic multipoles (axial tensor) 
and $T_{lm}$ in the second term the toroidal multipoles (polar tensor). 
Thus, the even-rank tensor $M_{lm}$ and the odd-rank tensor $T_{lm}$ become active only when both spatial-inversion and time-reversal symmetries are broken. 
The toroidal moment $\bm{t}$ appears in the lowest-rank ($l=1$) contribution in the latter term~\cite{LandauLifshitz198001,dubovik1990toroid,gorbatsevich1994toroidal}, which is written in the form 
\begin{eqnarray}
\bm{t} = 
\frac{1}{6c}\sum_{i}\bm{r}_{i}\times(\bm{r}_{i}\times\bm{j}_{i}).
\label{Eq:toroidal_charge}
\end{eqnarray}
Here, $\bm{r}_{i}$ and $\bm{j}_{i}$ are the position vector and the classical circular electric current 
at $\bm{r}_i$, respectively. 
When there is an internal magnetic field by spins instead of 
the electric current, the toroidal moment is expressed in terms of the localized spin $\bm{S}_{i}$ as
\begin{eqnarray}
\bm{t} =\frac{g \mu_{\rm{B}}}{2} \sum_{i} \bm{r}_{i} \times \bm{S}_{i}, 
\label{Eq:toroidal_spin}
\end{eqnarray}
where $g$ is the Land\'e $g$-factor and $\mu_{{\rm B}}$ is the Bohr magneton. 
This definition indicates that the toroidal moment is represented by the sum of 
vector products of the position vector and localized spin, 
as schematically shown in Fig.~\ref{Fig:Toroidal_structure}. 
We note that there is an ambiguity in the definitions in Eqs.~(\ref{Eq:toroidal_charge}) and~(\ref{Eq:toroidal_spin}) depending on the choice of the origin for $\bm{r}_{i}$~\cite{Spaldin:0953-8984-20-43-434203}.  

The toroidal moment $\bm{t}$ appears in 
the Hamiltonian under an inhomogeneous magnetic field $\bm{H}(\bm{r})$: 
the Hamiltonian can be expanded at some point $\bm{r}=0$ in terms of field gradients as 
\begin{eqnarray}
\mathcal{H}_{\rm ext}
= &-&\bm{m} \cdot \bm{H}(0) - \bm{t} \cdot [\nabla \times \bm{H}]_{\bm{r}=0}  \nonumber \\ 
&-&q_{\mu \nu}(\partial_{\mu}H_{\nu} +\partial_{\nu}H_{\mu})_{\bm{r}=0} + \cdots, 
\label{eq:H_int}
\end{eqnarray}
where $\bm{m}$ and $q_{\mu \nu}$ are the magnetic dipole and quadrupole moments, respectively 
($\mu, \nu=x,y,z$). 
The repeated greek indices are implicitly summed over hereafter. 
Equation~(\ref{eq:H_int}) indicates that the toroidal moment $\bm{t}$ couples to the curl of magnetic field, that is, the electric current. 

Next, we describe macroscopic responses of the toroidal moment 
to electromagnetic fields~\cite{LandauLifshitz198001,Spaldin2005renaissance}. 
Let us consider an expansion of the free energy with respect to the electric field $\bm{E}$ and the magnetic field $\bm{H}$ up to the second order, which is given by
\begin{align}
F(\bm{E}, \bm{H}) = F_0 - \frac{\varepsilon_{\mu \nu} E_{\mu}E_{\nu}}{8\pi} -\frac{\mu_{\mu \nu} H_{\mu} H_{\nu}}{8 \pi} -\alpha_{\mu \nu} E_{\mu} H_{\nu}. 
\label{Eq:free energy}
\end{align}
Here, $\varepsilon_{\mu \nu}$, $\mu_{\mu \nu}$, and $\alpha_{\mu \nu}$ are the dielectric permittivity, 
magnetic permeability, and magnetoelectric tensor, respectively. 
$\alpha_{\mu\nu}$ in the last term is related to linear magnetoelectric responses, and it is nonzero only when both global spatial-inversion and time-reversal symmetries are broken. 
This magnetoelectric contribution can be divided into three terms: 
\begin{eqnarray}
-a (\bm{E} \cdot \bm{H}) - \bm{T} \cdot (\bm{E} \times \bm{H}) - Q_{
\mu\nu} (E_{\mu} 
H_{\nu} 
+ E_{\nu} 
H_{\mu} 
), 
\label{Eq:ME free energy}
\end{eqnarray}
where coefficients of the first, second, and third terms represent magnetic flux (pseudoscalar), 
toroidal magnetization $\bm{T}$ (polar vector), and magnetic quadrupole (symmetric traceless pseudotensor). 
Thus, the antisymmetric components of magnetoelectric tensor $\alpha_{\mu\nu}$ correspond to the toroidal magnetization, which is defined as the toroidal moment per unit volume. 

The linear magnetoelectric effect by toroidal ordering is understood from Eq.~(\ref{Eq:ME free energy})~\cite{EdererPhysRevB.76.214404}. 
Namely, the second term implies the relations 
\begin{eqnarray}
\bm{P} \propto -\bm{T} \times \bm{H}, \quad  
\bm{M} \propto \bm{T} \times \bm{E}, 
\label{eq:PM}
\end{eqnarray}
where $\bm{P}$ and $\bm{M}$ are the electric polarization and magnetization, respectively. 
These indicate that the electric polarization (magnetization) is induced in the direction perpendicular to both of the toroidal magnetization and the magnetic (electric) field.

\section{Toroidal ordering in metals}
\label{Sec:Toroidal ordering in metals}

Here, we examine the nature of toroidal ordered states in crystals.  
First, we introduce lattice structures with local inversion symmetry breaking in Sec.~\ref{Sec:Lattices with local inversion symmetry breaking}. 
Next, we present a minimal low-energy Hamiltonian including the effect of the site-dependent antisymmetric spin-orbit coupling in Sec.~\ref{Sec:Model}. 
By analyzing the effective Hamiltonian on a layered honeycomb lattice, we investigate the influence of toroidal ordering on the electronic structure in Sec.~\ref{Sec:Electronic structure}, magnetotransport in Sec.~\ref{Sec:Magnetotransport}, and magnetoelectric effect in Sec.~\ref{Sec:Magnetoelectric effect}. 

\subsection{Lattices with local inversion symmetry breaking}
\label{Sec:Lattices with local inversion symmetry breaking}

\begin{figure}[t]
\begin{center}
\includegraphics[width=1.0 \hsize]{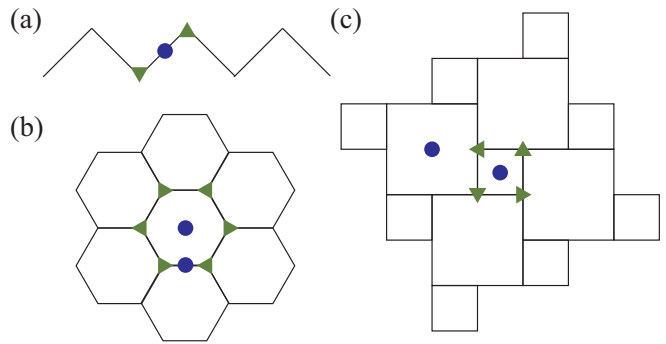} 
\caption{
\label{Fig:localinversionbreaking}
(Color online) 
Schematic pictures of lattice structures with local inversion symmetry breaking: 
(a) one-dimensional zigzag chain, (b) honeycomb lattice, and (c) 1/5-depleted square lattice. 
Circles indicate the inversion centers, while triangles represent the lattice sites at which the inversion symmetry is broken. 
}
\end{center}
\end{figure}

A minimal ingredient to activate a toroidal moment is spatial-inversion symmetry breaking at each magnetic site. 
In such situation, an odd-parity crystalline electric field is present at the magnetic sites, which mixes orbitals with different parities. 
Such local parity mixing together with an atomic spin-orbit coupling plays an important role in realizing 
toroidal ordering in metals, as we will see in the following sections. 

In order to demonstrate the above scenario, we focus on the lattice structures in which the inversion symmetry is broken locally at each site but the global inversion symmetry is preserved at off-site positions. 
There are several lattices with such local inversion symmetry breaking; 
for instance, a one-dimensional (1D) zigzag chain [Fig.~\ref{Fig:localinversionbreaking}(a)], 
two-dimensional honeycomb lattice [Fig.~\ref{Fig:localinversionbreaking}(b)], 1/5-depleted square lattice [Fig.~\ref{Fig:localinversionbreaking}(c)], and 
a three-dimensional (3D) diamond lattice. 
On these lattices, the spatial-inversion symmetry is broken at the lattice sites (triangles in Fig.~\ref{Fig:localinversionbreaking}), although it is preserved at the centers of bonds or plaquettes (circles).

\subsection{Effective single-band model}
\label{Sec:Model}

In order to describe origins of essential ingredients for a minimal tight-binding model, 
we start from a generic four-band model with local parity mixing. 
Following the procedure given in Appendix~\ref{sec:Appendix1}, 
we obtain the effective single-band model, which will be examined mainly in the following sections. 

The Hamiltonian for the generic four-band model is given by
\begin{align}
\mathcal{H}_{\rm 4-band} = \mathcal{H}_{\rm kin} + \mathcal{H}_{\rm hyb} + \mathcal{H}_{\rm o-CEF} + \mathcal{H}_{\rm LS} + \mathcal{H}_{\rm int},
\label{eq:4band_H}
\end{align}
where
\begin{align}
&\mathcal{H}_{{\rm kin}} = -\sum_{
i,j
}
\sum_{\alpha=s,p_x,p_y,p_z} \sum_{\sigma}
(\tilde{t}^{\alpha}_{ij} 
\tilde{c}_{i \alpha \sigma}^{\dagger} \tilde{c}_{j \alpha \sigma} + {\rm H.c.}), 
\label{eq:H_kin} 
\\
&\mathcal{H}_{{\rm hyb}} = -\sum_{\langle i,j \rangle}
\sum_{\alpha 
=p_x,p_y,p_z 
} \sum_{\sigma}
(\tilde{V}^{\alpha}_{ij} 
\tilde{c}_{i s 
\sigma}^{\dagger} \tilde{c}_{j 
\alpha 
\sigma} + {\rm H.c.}), 
\label{eq:H_hyb}
\\
&\mathcal{H}_{{\rm o-CEF}} = \sum_{i}
\sum_{\alpha=p_x,p_y,p_z} \sum_{\sigma}
\tilde{D}_{i}^{\alpha} ( \tilde{c}_{i s \sigma}^{\dagger} \tilde{c}_{i \alpha \sigma} + {\rm H.c.}), 
\label{eq:H_CEF}
\\
&\mathcal{H}_{{\rm LS}} = \frac{\lambda}{2}\sum_{i}\sum_{\alpha,\beta=p_x,p_y,p_z}  \sum_{\sigma, \sigma'}
\tilde{c}_{i \alpha \sigma}^{\dagger} \tilde{H}_{{\rm LS}} \tilde{c}_{i \beta \sigma'}, 
\label{eq:H_LS}
\\
&\mathcal{H}_{\rm int} = 
\sum_i \sum_{\alpha\beta\alpha'\beta'} \sum_{\sigma\sigma'} U_{\alpha\beta\alpha'\beta'} \tilde{c}_{i\alpha\sigma}^\dagger \tilde{c}_{i\beta\sigma'}^\dagger \tilde{c}_{i\beta'\sigma'} \tilde{c}_{i\alpha'\sigma}. 
\label{eq:H_int_Hubbard}
\end{align}
Here, $\tilde{c}_{i \alpha \sigma}^{\dagger}$ ($\tilde{c}_{i \alpha \sigma}$) is the creation (annihilation) operator of a conduction electron with orbital $\alpha$ and spin $\sigma$ at site $i$; we consider four orbitals, i.e., ``$s$-type" orbital with the angular momentum $l=0$ (even parity) and three ``$p$-type" 
orbitals with $l=1$ (odd parity). 
This is a minimum set of orbitals for describing the parity mixing. 
Equation~(\ref{eq:H_kin}) represents the kinetic energy of conduction electrons; 
the sum is limited to on-site and nearest-neighbor sites. 
The on-site part describes the atomic energy, which is set as $\tilde{t}_{ii}^s = \tilde{E}^s$ and $\tilde{t}_{ii}^{\alpha=p_x,p_y,p_z} = 0$. 
Equation~(\ref{eq:H_hyb}) describes the off-site hybridization between $s$ and $p$ orbitals; the sum $\langle i,j \rangle$ is taken for the nearest-neighbor sites, and $\tilde{V}_{ij}^\alpha$ depends 
on the $p$ orbital as well as the bond direction. 
Equation~(\ref{eq:H_CEF}) 
denotes the 
odd-parity crystalline electric field; $\tilde{D}_i^\alpha$ depends on both site and orbital. 
Note that $\tilde{D}_i^\alpha$ is nonzero only in the absence of local inversion symmetry, while $\tilde{V}_{ij}^{\alpha}$ is always present even with local inversion symmetry. 
We note that both $\tilde{V}_{ij}^\alpha$ and $\tilde{D}_i^\alpha$ are indispensable 
to obtain the site-dependent antisymmetric spin-orbit coupling in the effective model (see Appendix~\ref{sec:Appendix1}). 
Equation~(\ref{eq:H_LS}) represents the atomic spin-orbit coupling for $p$ orbitals with $l=1$: 
$\tilde{H}_{{\rm LS}}$ is the $6 \times 6$ matrix given by 
\begin{eqnarray}
\tilde{H}_{{\rm LS}}= 
\left( 
\begin{array}{ccc}
0 & -{\rm i} \sigma^{z} & {\rm i} \sigma^y   \\
{\rm i} \sigma^z & 0 & -{\rm i} \sigma^x \\
- {\rm i} \sigma^y &{\rm i} \sigma^x & 0  \\
\end{array} \right),
\end{eqnarray}
where 
$
\sigma^{
\mu}$ is the 
$\mu$ component of Pauli matrix for spin. 
Finally, Eq.~(\ref{eq:H_int_Hubbard}) is a general form of on-site Coulomb interactions. 
Here, we only take account of the intraorbital components
for simplicity, i.e., $U_{\alpha
\beta
\alpha'
\beta'} = (U/2) \delta_{\alpha\beta} \delta_{\alpha'\beta'} \delta_{\alpha\alpha'}$ ($\delta_{\alpha\beta}$ is the Kronecker delta). 

We treat the interaction term at the level of a mean-field approximation to allow magnetic solutions. 
The mean-field form is given by
\begin{align}
\mathcal{H}_{\rm int}^{\rm MF} = -
\sum_{i}\sum_{\alpha=s,p_x,p_y,p_z} 
\tilde{\bm{M}}_{i}^{\alpha} \cdot 
\tilde{\bm{s}}_i^\alpha,
\label{eq:H_int_MF}
\end{align}
where the mean field $\tilde{\bm{M}}^\alpha_i = 
2 U \tilde{\bm{m}}^\alpha_i $ 
and the magnetic moment $\tilde{\bm{m}}^\alpha_i = 
\langle \tilde{\bm{s}}_i^\alpha \rangle =
\langle 
\sum_{\sigma \sigma'} 
\tilde{c}_{i \alpha \sigma}^{\dagger} (\bm{\sigma}_{\sigma \sigma'}/2) \tilde{c}_{i \alpha \sigma'} \rangle$. 
We omit a constant from the mean-field decoupling here and in Sec.~\ref{Sec:Toroidal ordering in metals}.

In the present study, we focus on a 3D system composed of weakly-coupled uniform 1D chains running in the $z$ direction. 
By several simplifications in the limit of strong spin-orbit coupling (see Appendix~\ref{sec:Appendix1}), 
we arrive at an effective single-band model, whose Hamiltonian is given by 
\begin{align}
\mathcal{H}^{\rm MF} 
=& -t \sum_{\langle i,j \rangle}
 \sum_{\sigma}
( c_{i \sigma}^{\dagger}c_{j \sigma} + {\rm H.c.}) \nonumber \\ 
& +
2\sum_{i} 
(\bm{s}_{i}\times \bm{D}_i)^{z} 
- \sum_{i} \bm{M}_{i} \cdot \bm{s}_i. 
\label{Hubbard_Ham_mean}
\end{align}
Here, $c_{i \sigma}^{\dagger}$ ($c_{i \sigma}$) is the creation (annihilation) operator of a conduction electron in the effective single band at site $i$ and quasispin $\sigma$, 
which distinguishes the time-reversal pair states; 
$\bm{s}_i = \sum_{\sigma,\sigma'} c_{i \sigma}^{\dagger} (\bm{\sigma}_{\sigma \sigma'}/2) c_{i \sigma'}$. 
The first term is the kinetic energy of electrons with a renormalized hopping $t$; 
we assume the isotropic hopping for the in-chain and out-of-chain directions for simplicity, but the anisotropic case gives qualitatively the same results. 
Hereafter, we set $t=1$. 
The second term represents the site-dependent antisymmetric spin-orbit coupling, in which 
$\bm{D}_{i}$ is a site-dependent antisymmetric vector originating from 
the odd-parity crystalline electric field $\tilde{D_i^{\alpha}}$, off-site hybridization $\tilde{V}_{ij}^{\alpha}$, and atomic spin-orbit coupling $\lambda$ 
[see Eq.~(\ref{eq:DVp}) in Appendix~\ref{sec:Appendix1}]. 
The third term in Eq.~(\ref{Hubbard_Ham_mean}) describes the mean-field form of the 
Coulomb interaction between electrons; 
$\bm{M}_i = 
2U\bm{m}_i 
$, where $\bm{m}_i = 
\langle 
\bm{s}_i \rangle$ is the magnetic moment at site $i$.

\begin{figure}[t]
\begin{center}
\includegraphics[width=1. \hsize]{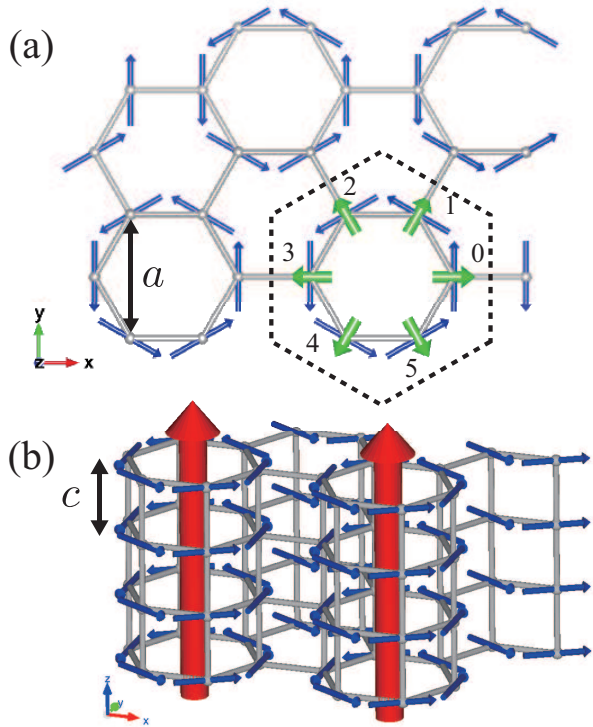} 
\caption{
(Color online) 
(a) Schematic picture of a projection of the layered honeycomb lattice onto the $xy$ plane. 
The thin (blue) arrows show the magnetic pattern assumed 
in the toroidal ordered state [Eq.~(\ref{eq:M})], and the thick (green) arrows indicate the specific 
directions of the odd-parity crystalline electric field [Eq.~(\ref{eq:D})]. 
The dashed hexagon represents the six-sublattice unit cell with the sublattice indices ($l=0$, $1$, $\cdots$, $5$). 
$a$ is the lattice constant in the $xy$ plane. 
(b) Schematic picture of the layered honeycomb lattice. 
The red arrows in the $z$ direction represent the toroidal magnetizations. 
$c$ is the lattice constant in the $z$ direction. 
\label{Fig:lattice}
}
\end{center}
\end{figure}

In the following, we consider the model in Eq.~(\ref{Hubbard_Ham_mean}) on a stacked honeycomb lattice, as shown in Fig.~\ref{Fig:lattice}. 
We take the lattice constants $a=c=1$. 
Extensions to other lattices with local inversion symmetry breaking are straightforward. 
For the present stacked honeycomb-lattice case, we assume the presence of $\bm{D}_i$ in a six-sublattice form with the specific directions in the $xy$ plane, 
as shown by the thick arrows in Fig.~\ref{Fig:lattice}(a). 
We also assume that the same patterns of $\bm{D}_i$ are stacked along the $z$ direction. 
The form in the six-site sublattice is represented by 
\begin{align}
\bm{D}_{l} 
= D \sin k_z 
\left( \cos 
\frac{\pi}{3} l, 
\, \sin 
\frac{\pi}{3} l, 
\, 0 \right), 
\label{eq:D}
\end{align}
where $l$ is the sublattice index ($
l=0$, $1$, $\cdots$, $5$) 
and $D$ is a parameter to control the magnitude of the antisymmetric spin-orbit coupling. 
It is worthy noting that the factor of $\sin k_z$ always enters into $\bm{D}_{l}$ 
in the quasi-1D systems; see the derivation in Appendix~\ref{sec:Appendix1}. 

Furthermore, in the following sections, 
we assume a six-sublattice vortex-type magnetic order as shown in Fig.~\ref{Fig:lattice}(a) in a mean-field form. 
We call it as the toroidal magnetic order hereafter. 
Specifically, the mean field for the toroidal magnetic order is given by 
\begin{align}
\bm{M}_l = M_{\bm{T}} \left( -\sin \frac{\pi}{3} l, \, \cos \frac{\pi}{3} l, \, 0 \right). 
\label{eq:M}
\end{align}
$M_{\bm{T}}$ 
is a measure of the mean field for the toroidal magnetic order, which 
is treated as a free parameter throughout in this section, 
while it will be determined by solving the self-consistent equations in 
Sec.~\ref{Sec:Mean field calculations}. 
We note that a similar magnetic pattern was 
indeed observed in the partial ordered state below 20~K in an Uranium compound 
UNi$_4$B~\cite{Mentink1994}. 
We will remark on this point in Sec.~\ref{Sec:Summary}. 

Since $\bm{D}_l$ plays a role of a local electric field and $\bm{D}_l \perp \bm{M}_l$ at each site, 
the toroidal magnetic order $\bm{M}_{l}$ must accompany a ferroic order of toroidal moments,
$\bm{T}\propto\bm{D}_{l}\times\bm{M}_{l}=(0,0,DM_{\bm{T}}\sin k_{z})$, as shown in Fig.~\ref{Fig:lattice}(b). 
In other words, even if a macroscopic polarization is absent, a spontaneous toroidal order can be realized by specific magnetic ordering with underlying local inversion symmetry breaking.

\subsection{Electronic structure}
\label{Sec:Electronic structure}

\begin{figure}[t]
\begin{center}
\includegraphics[width=1.0 \hsize]{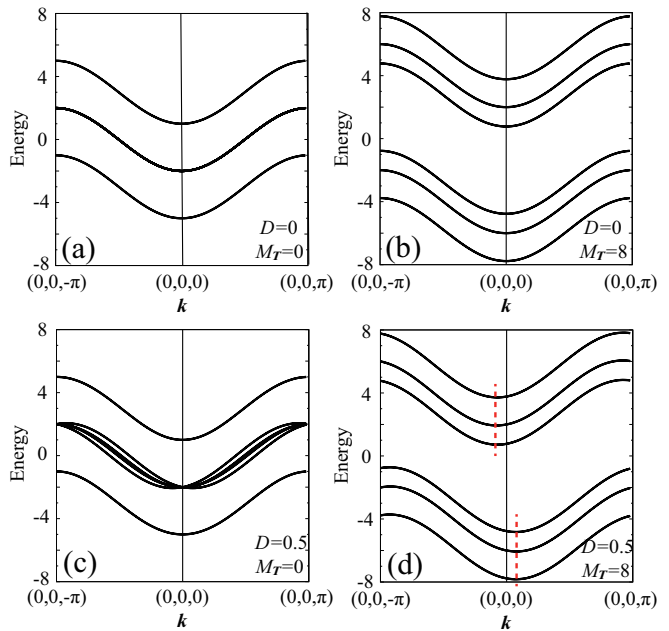} 
\caption{
(Color online)
Energy dispersion along the $k_z$ direction of the Hamiltonian in Eq.~(\ref{Hubbard_Ham_mean}) with Eqs.~(\ref{eq:D}) and (\ref{eq:M})
for (a) $D=0$, $M_{\bm{T}}=0$, (b) $D=0$, $M_{\bm{T}}=8$, (c) $D=0.5$, $M_{\bm{T}}=0$, and (d) $D=0.5$, $M_{\bm{T}}=8$. 
The dashed red lines in (d) indicate the band bottoms in the $k_z$ direction. 
\label{Fig:Honeycomb_banddispersion}
}
\end{center}
\end{figure}

Let us first discuss the electronic structure of the model in Eq.~(\ref{Hubbard_Ham_mean}) with Eqs.~(\ref{eq:D}) and (\ref{eq:M}). 
Figure~\ref{Fig:Honeycomb_banddispersion} shows the band structures in the $k_z$ direction from 
$\bm{k}=(0,0,-\pi)$ to $(0,0,\pi)$ for several values of $D$ and $M_{\bm{T}}$. 

Figure~\ref{Fig:Honeycomb_banddispersion}(a) shows the band structure in the paramagnetic state ($M_{\bm{T}}=0$) at $D=0$, where both spatial-inversion and time-reversal symmetries are preserved. 
In this case, there are three bands; the top and bottom bands are doubly degenerate each, while the middle one is eightfold degenerate. 
When only the time-reversal symmetry is broken by 
the toroidal magnetic order $M_{\bm{T}} \neq 0$ at $D=0$, 
the bands are split into two bunches depending on the spins parallel or antiparallel to the mean field $
\bm{M}_l$.
Each bunch consists of three bands, each of which remains to be doubly degenerate due to the global inversion symmetry: 
the band dispersions satisfy the relation 
$\epsilon_\sigma(\bm{k}) = \epsilon_\sigma(-\bm{k})$. 
Meanwhile, when $D$ is nonzero representing local inversion symmetry breaking and the system is 
in the paramagnetic state ($M_{\bm{T}}=0$), 
the antisymmetric spin splitting of the bands occurs, 
as shown in Fig.~\ref{Fig:Honeycomb_banddispersion}(c). 
Here, the time reversal symmetry ensures $\epsilon_\sigma(\bm{k}) = \epsilon_{-\sigma}(-\bm{k})$.  
It should be noted however that each band is still doubly degenerate as the global inversion symmetry remains: $\epsilon_\sigma(\bm{k}) = \epsilon_\sigma(-\bm{k})$. 
This is different from the case with an ordinary antisymmetric spin-orbit coupling such as the Rashba-type one where  
{\it global} inversion symmetry is broken. 

What happens when the system exhibits a toroidal order, i.e., for $M_{\bm{T}} \neq 0$ and $D \neq 0$ ? 
In this case, both spatial-inversion $\mathcal{P}$ and time-reversal $\mathcal{T}$ symmetries are broken. 
Each band, however, retains twofold degeneracy owing to the combined $\mathcal{PT}$ symmetry, which ensures 
$\epsilon_\sigma(\bm{k}) = \epsilon_{-\sigma}(\bm{k})$. 
However, there is no guarantee that the energy at $\bm{k}$ is degenerate with that at $-\bm{k}$. 
In fact, the resultant band structure consists of six bands, doubly degenerate each, with a shift of the band bottom 
from the $\Gamma$ point, as shown in Fig.~\ref{Fig:Honeycomb_banddispersion}(d).  

The modulation of the bands with a shift of the band bottom is understood in the single-chain limit as follows. 
The Hamiltonian for the single chain is given by a simple form of the $2 \times 2$ matrix: 
\begin{align}
{\mathcal{H}_{\rm 1D}^{\rm MF}} 
= 
\left( 
\begin{array}{cc}
-2t \cos k_z & {\rm i} D^{-} +M^{-}/2   \\
-{\rm i} D^{+} +M^{+}/2  & -2t \cos k_z 
\end{array} \right),  
\label{eq:localeff}
\end{align}
where $D^{\pm} = D^{x}  \pm {\rm i} D^{y}$ and $M^{\pm} = M^{x}  \pm {\rm i} M^{y}$. 
Here, $\bm{D} = (D^x, D^y, 0)$ and $\bm{M} = (M^x, M^y, 0)$. 
See also Eq.~(\ref{eq:H_1D_x}) in Appendix~A. 
The energy spectrum of this Hamiltonian is given by
\begin{align}
\epsilon (k_z) = -2 t \cos k_z \pm \sqrt{\bm{D}^2  + \bm{M}^2/4 -(\bm{D} \times \bm{M})^{z}}.
\label{eq:dispersion_1chain}
\end{align}
As $\bm{D}$ is proportional to $\sin k_z$ [see Eq.~(\ref{eq:D})] and the last term in the square roots is linear in $\bm{D}$, a shift of the band bottom occurs 
when $\bm{D}\times\bm{M}\neq0$, which is proportional to the toroidal magnetization. 

The results indicate that the toroidal magnetization acts as an effective gauge field for conduction electrons. 
The shifted band structure, however, does not generate spontaneous electric current in the equilibrium state due to the gauge invariance~\cite{volkov1981macroscopic,Yanase:JPSJ.83.014703}. 
The asymmetric band structure can be detected in principle by experiments, such as the angle-resolved photo-emission spectroscopy. 
We expect that the peculiar band deformation may become the origin of the nonlinear optical effect, such as a nonreciprocal directional dichroism~\cite{li2013coupling}. 

We note that the system has a particle-hole symmetry. 
Indeed, the Hamiltonian in Eq.~(\ref{Hubbard_Ham_mean}) is unchanged by the particle-hole transformation, 
($c_{l\sigma}$, $c_{l\sigma}^\dagger$) $\to$ $(-1)^l$ ($d_{l\sigma}^\dagger$, $d_{l\sigma}$) with the shift of $k_z$ by $\pi$ and $\bm{M}_l \to -\bm{M}_l$. 
This symmetry is also seen in the band structure in Fig.~\ref{Fig:Honeycomb_banddispersion}; the band dispersions satisfy $\epsilon_{\sigma}(k_z) = -\epsilon_{-\sigma}(\pi-k_z)$. 

\subsection{Magnetotransport}
\label{Sec:Magnetotransport}

\begin{figure}[t]
\begin{center}
\includegraphics[width=1.0 \hsize]{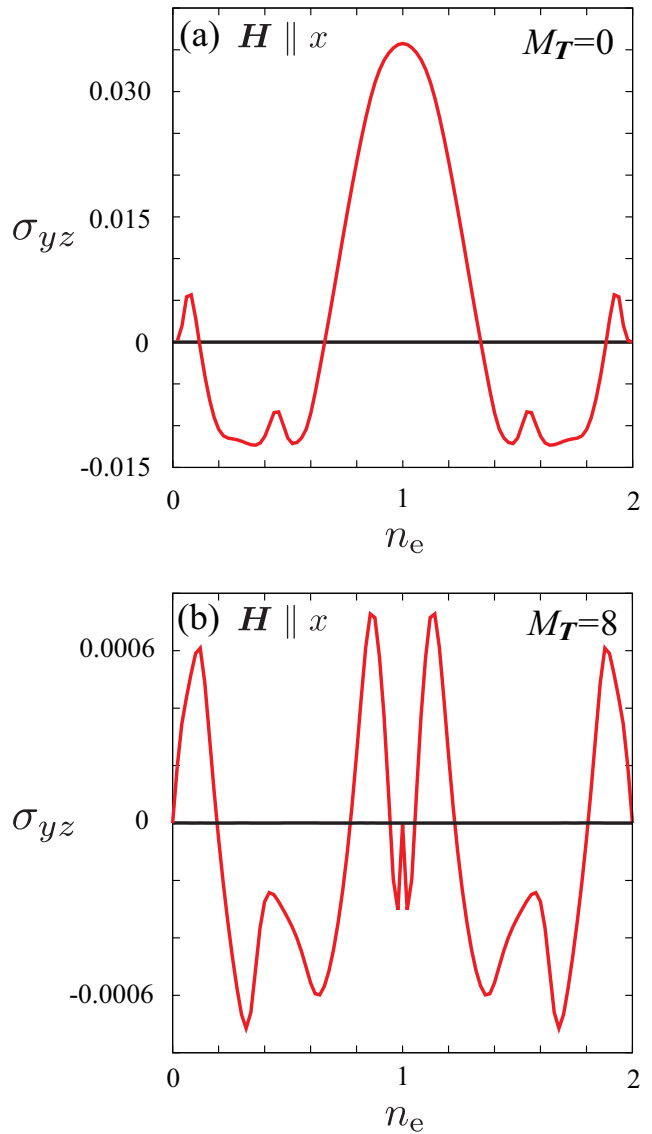} 
\caption{
(Color online) 
Electron density dependence of the Hall conductivity for (a) $M_{\bm{T}}=0$ and (b) $M_{\bm{T}}=8$. 
The data are obtained at $D=0.5$ in the magnetic field $H=0.5$ applied in the $x$ direction. 
We take the broadening factor $\delta=0.01$ and temperature $T=0.1$. 
\label{Fig:toroidal_sigmayz}
}
\end{center}
\end{figure}

\begin{figure*}[t]
\begin{center}
\includegraphics[width=1.0 \hsize]{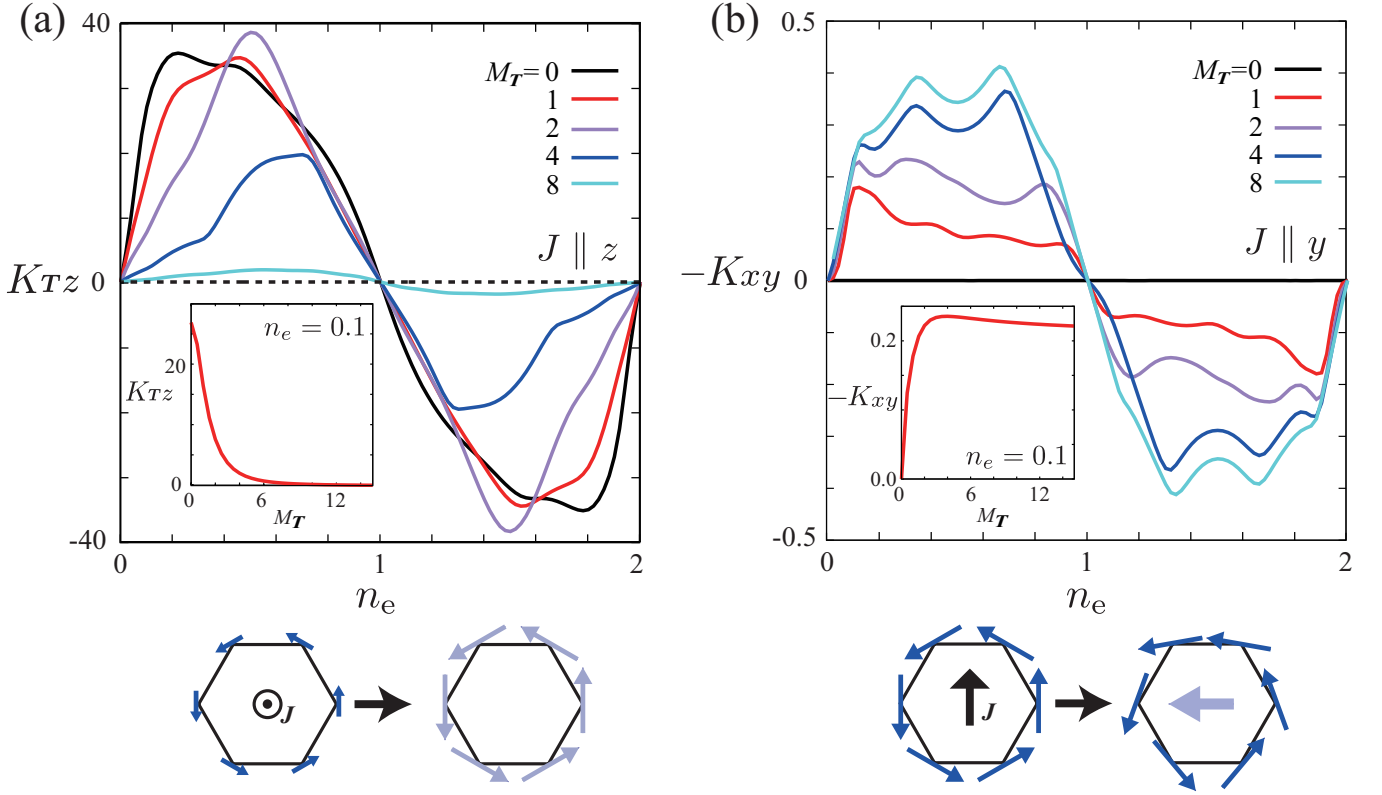} 
\caption{
(Color online) 
Electron density dependence of 
(a) the toroidal-current correlation $K_{\bm{T}z}$ and (b) the magnetization-current correlation $K_{xy}$. 
The insets of (a) and (b) show their $M_{\bm{T}}$ dependences at $n_{{\rm e}}=0.1$. 
The data are obtained for $D=0.5$, $\delta=0.01$ and $T=0.1$. 
Schematic pictures for the magnetoelectric responses are shown in the bottom panels. 
In each picture, the left and right pictures show the alignment of the magnetic moments before and after the electric current is applied. 
In the rightmost one, the arrow in the center of hexagon indicates a net uniform magnetization induced by the electric current. 
\label{Fig:SAFJZ}
}
\end{center}
\end{figure*}

We here discuss magnetotransport coefficients for the model in Eq.~(\ref{Hubbard_Ham_mean}) with Eqs.~(\ref{eq:D}) and (\ref{eq:M}). 
We calculate the conductivity tensor in terms of the current-current correlation by the standard Kubo formula as
\begin{align}
\sigma_{\mu \nu} =\frac{e^2}{\hbar}\frac{1}{{\rm i} V} \sum_{m,n,\bm{k}} \frac{f(\epsilon_{n \bm{k}})-f(\epsilon_{m \bm{k}})}{\epsilon_{n \bm{k}}-\epsilon_{m \bm{k}}} 
\frac{J_{\mu,\bm{k}}^{nm} J_{\nu,\bm{k}}^{mn}
}{
\epsilon_{n \bm{k}}-\epsilon_{m \bm{k}}+{\rm i} \delta}, 
\end{align}
where $V$ is the system volume, $f(\epsilon)$ is the Fermi distribution function, $J_{\mu,\bm{k}}^{nm}=\langle n\bm{k} | J_{\mu} | m \bm{k} \rangle$ 
[$J_{\mu}$ is 
the current operator in the direction $\mu=(x,y,z)$], and 
$\epsilon_{m \bm{k}}$ and $|m \bm{k} \rangle$ are the eigenvalue and eigenstate of ${\cal H}^{\rm MF}$ with the Zeeman term (see below). 
Here, $m$ and $n$ are the band indices. 
We take $e^2/h=1$ ($e$ is the elementary charge and $h$ is the Planck constant), a broadening factor $\delta = 0.01$ and temperature $T=0.1$. 
The summation of $\bm{k}$ is taken over the folded Brillouin zone in magnetically ordered state. 

Figure~\ref{Fig:toroidal_sigmayz} shows 
the Hall conductivity $\sigma_{yz}$ as a function of the electron density $n_{{\rm e}}= (1/N) \sum_{i \sigma} \langle c_{i \sigma}^{\dagger} c_{i \sigma}\rangle$, where $N$ is the total number of sites. 
The results are obtained for $D=0.5$ in a magnetic field applied in the $x$ direction; 
here, we added the Zeeman term $\mathcal{H}_{\rm Z} = -H\sum_i s_i^x$ to the Hamiltonian $\mathcal{H}^{\rm MF} $ in Eq.~(\ref{Hubbard_Ham_mean}). 
The results are symmetric with respect to $n_{\rm e}=1$ because of the particle-hole symmetry discussed in the end of the previous section. 

Figure~\ref{Fig:toroidal_sigmayz}(a) shows a Hall conductivity in the absence of the toroidal magnetic order $M_{\bm{T}}=0$.
The sign and magnitude of the Hall conductivity depends on the electron density $n_{{\rm e}}$ reflecting a nature of carriers near the Fermi level.
It is highly anisotropic; only the $\sigma_{yz(xz)}$ component becomes nonzero in the magnetic field applied in the $x$($y$) direction.
This is a consequence of the presence of the (site-dependent) antisymmetric spin-orbit coupling.
Indeed, $\sigma_{\mu\nu}$ disappears if we take $D=0$.
The toroidal magnetic order affects significantly the behavior of the Hall conductivity as shown in Fig.~\ref{Fig:toroidal_sigmayz}(b) for $M_{\bm{T}}=8$.
Although the anisotropy of $\sigma_{\mu\nu}$ does not change in the presence of $M_{\bm{T}}$, the Hall conductivity is strongly suppressed.
As will be shown in the next section, $M_{\bm{T}}$ has strong temperature dependence in the ordered state, and hence $\sigma_{\mu\nu}$ should exhibit strong suppression with decrease of temperature.

\subsection{Magnetoelectric effect}
\label{Sec:Magnetoelectric effect}

Now let us discuss magnetoelectric effects in the toroidal ordered state. 
We compute the linear response function in terms of the correlation between the (toroidal) magnetization and an electric current by electric field in the form 
\begin{align}
K_{\mu \nu} =\frac{g \mu_{{\rm B}}}{2}\frac{e}{\hbar}\frac{1}{{\rm i} V} \sum_{m,n,\bm{k}} \frac{f(\epsilon_{n \bm{k}})-f(\epsilon_{m \bm{k}})}{\epsilon_{n \bm{k}}-\epsilon_{m \bm{k}}} 
\frac{
\sigma_{\mu,\bm{k}}^{nm} J_{\nu,\bm{k}}^{mn} 
}{
\epsilon_{n \bm{k}}-\epsilon_{m \bm{k}}+{\rm i} \delta}, 
\label{Eq:Mangetoelectric effect}
\end{align}
where 
$\sigma_{\mu,\bm{k}}^{nm}=\langle n\bm{k} | \sigma_{\mu} | m \bm{k} \rangle$.  
We take $g \mu_{{\rm B}} e/2h=1$, 
$D=0.5$, $\delta=0.01$, and $T=0.1$. 
In the following, we discuss two different types of magnetoelectric effects: 
one is the toroidal magnetic response ($\mu=\bm{T}$) to an electric current, and the other is the uniform magnetization ($\mu=x,y,z$) induced by an electric current.

\subsubsection{Longitudinal toroidal magnetization by electric current}
\label{sec:ME1}

First, we focus on the toroidal magnetic response to an electric current. 
Here, we consider $\sigma_{\bm{T},\bm{k}}^{nm}$ in Eq.~(\ref{Eq:Mangetoelectric effect}) 
as the toroidal magnetic order in Eq.~(\ref{eq:M}). 
Hence, for instance, $K_{\bm{T}z}$ is the coefficient for the toroidal magnetic order 
induced by the electric current in the $z$ direction.
Among $K_{\bm{T}\mu}$, only $K_{\bm{T}z}$ becomes nonzero, as will be discussed 
in Sec.~\ref{sec:summary_of_ME} and the table in Fig.~\ref{Fig:table_response}.  

Figure~\ref{Fig:SAFJZ}(a) shows $K_{\bm{T}z}$ as a function of the electron density $n_{{\rm e}}$. 
$K_{\bm{T}z}$ becomes nonzero in the entire region of $n_{{\rm e}}$ except for the insulating cases at $n_{{\rm e}}=0$, 1, and 2. 
Note that $K_{{\bm T}z}$ is antisymmetric with respect to $n_{\rm e}=1$ because of the sign change of the magnetic moment 
in the particle-hole transformation discussed in Sec.~\ref{Sec:Electronic structure}. 
The result indicates that the toroidal magnetization can be induced by the electric current in the $z$ direction. 
This provides the possibility in experiments to align the toroidal domains by cooling the system  
in a current flow perpendicular to the planes. 

Although the toroidal magnetic response shows complicated behavior depending on both $n_{{\rm e}}$ and $M_{\bm{T}}$, it tends to be smaller for larger $M_{\bm{T}}$. 
This tendency is clearly seen 
in the low and high density regions where the Fermi surface has a simple shape. 
The inset of Fig.~\ref{Fig:SAFJZ}(a) displays the behavior of $K_{\bm{T}z}$ at low density $n_{\rm e}=0.1$ as a function of $M_{\bm{T}}$; 
the toroidal magnetic response is largest at 
$M_{\bm{T}}=0$, and is suppressed as $M_{\bm{T}}$ increases. 
This susceptibility-like behavior suggests that 
$K_{\bm{T}z}$ as a function of temperature becomes largest near the critical temperature for the toroidal ordered state, 
as long as the transition is of second order. 
Indeed, we will see such behavior in the mean-field calculation in Sec.~\ref{Sec:Finite temperature property}. 

We note that $K_{\bm{T}z}$ substantially depends on the broadening factor $\delta$ in Eq.~(\ref{Eq:Mangetoelectric effect}). 
This indicates that this quantity has a dominant contribution from the intraband 
components with $m=n$ in Eq.~(\ref{Eq:Mangetoelectric effect}).

\subsubsection{Transverse magnetization by electric current}
\label{sec:ME2}

Next, we discuss another magnetoelectric effect, the transverse magnetic response to an electric current. 
Here, we consider $\sigma_{\mu,\bm{k}}^{nm}$ with $\mu=x,y,z$ in Eq.~(\ref{Eq:Mangetoelectric effect}). 
For instance, $K_{xy}$ is the coefficient for the uniform magnetization in the $x$ direction induced by the electric current in the $y$ direction. 
Among $K_{\mu\nu}$, only the transverse components within the plane, i.e., $K_{xy}$ and $K_{yx}$ become nonzero, and they satisfy 
the antisymmetric relation $K_{xy}=-K_{yx}$ deduced from 
Eq.~(\ref{Eq:ME free energy}); see also the discussion for the table in Fig.~\ref{Fig:table_response} in Sec.~\ref{sec:summary_of_ME}. 

Figure~\ref{Fig:SAFJZ}(b) shows the result of $K_{xy}$ as a function of the electron density $n_{{\rm e}}$. 
Similar to $K_{\bm{T}z}$ in Fig.~\ref{Fig:SAFJZ}(a), $K_{xy}$ shows a nonzero value in the entire region of $n_{{\rm e}}$, except for the insulating cases at $n_{{\rm e}}=0$, 1, and 2, and it is antisymmetric with respect to $n_{\rm e}=1$. 
The result indicates that a uniform magnetization can be induced by an electric current when the toroidal order is present ($M_{\bm{T}}\neq 0$). 
This is considered as a multiferroic response in metals with simultaneous symmetry breaking of spatial-inversion and time-reversal. 
Experimentally, a nonzero value of $K_{xy}=-K_{yx}$ is an indication of the toroidal order. 

The magnitude of the induced magnetization becomes larger for larger $M_{\bm{T}}$, in contrast to the toroidal magnetic response $K_{\bm{T}z}$ in Fig.~\ref{Fig:SAFJZ}(a). 
$K_{xy}=0$ for $M_{\bm{T}}=0$ is due to the presence of time-reversal symmetry, and 
$K_{xy}$ increases as increase of $M_{\bm{T}}$ 
for small $M_{\bm{T}}$ and it almost saturates for large $M_{\bm{T}}$, as shown in the inset of Fig.~\ref{Fig:SAFJZ}(b). 
In contrast to $K_{\bm{T}z}$, $K_{xy}$ 
weakly depends on $\delta$, indicating that a dominant contribution comes from the interband components in Eq.~(\ref{Eq:Mangetoelectric effect}).

\subsubsection{Summary of magnetoelectric effects}
\label{sec:summary_of_ME}

\begin{figure}[t]
\begin{center}
\includegraphics[width=1.0 \hsize]{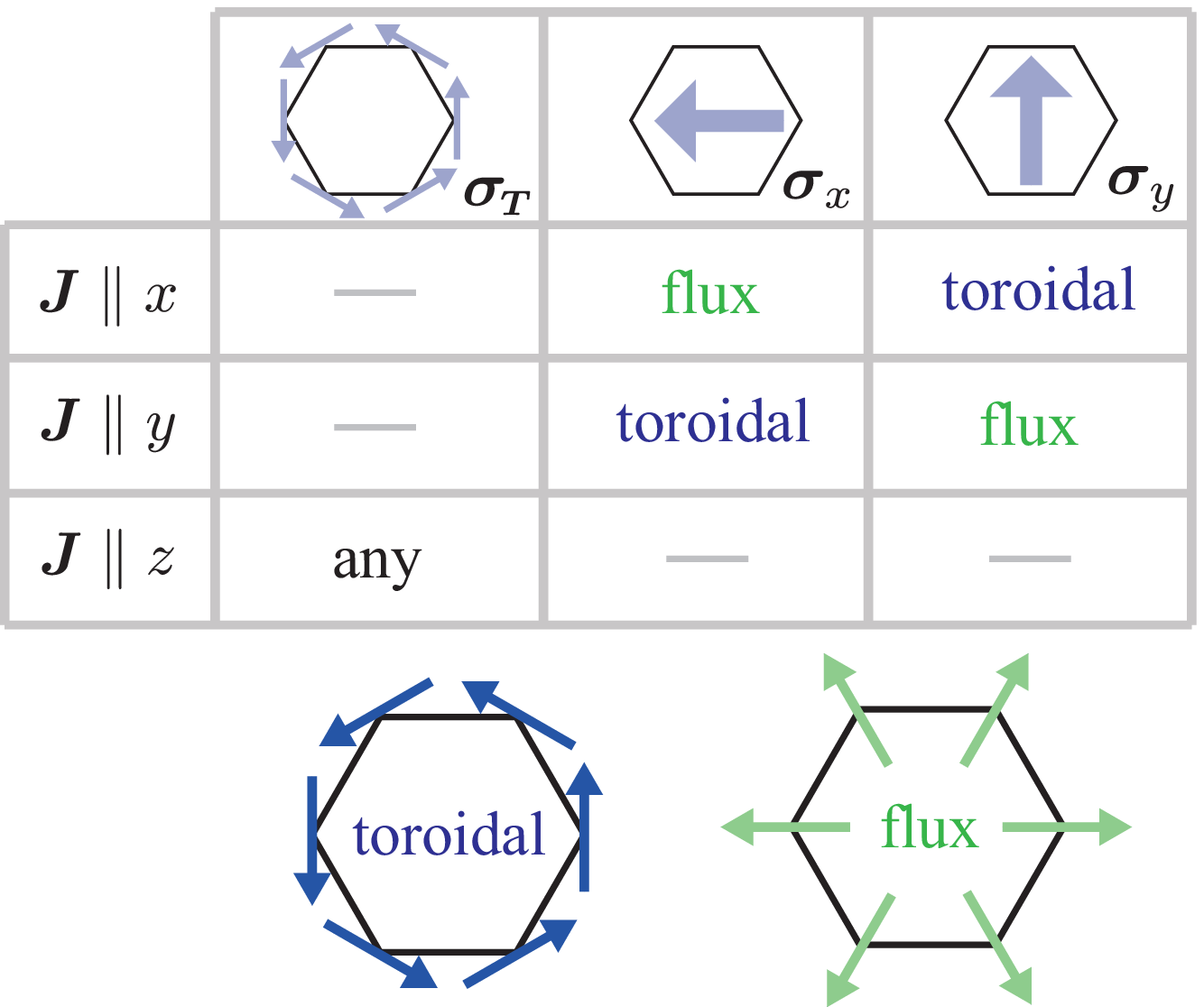} 
\caption{
\label{Fig:table_response}
(Color online) 
Summary of the magnetoelectric effects under toroidal or flux orders. 
In the table, ``toroidal" and ``flux" indicate that by which the magnetism in the top row is induced by the electric current in the left column. 
``Any" represents paramagnetic or any magnetic ordered states including the toroidal and flux orders.
Schematic pictures of the toroidal and flux orders are shown in the bottom; 
the arrows indicate the magnetic moments. 
}
\end{center}
\end{figure}

We summarize the results of magnetoelectric effects obtained for the model in Eq.~(\ref{Hubbard_Ham_mean}) with Eqs.~(\ref{eq:D}) and (\ref{eq:M}). 
Figure~\ref{Fig:table_response} shows the table for the magnetoelectric effects in terms of the applied electric current (in the left column) and the resultant magnetic response (in the top row). 
``Toroidal'' or ``flux'' in the table represents the underlying magnetic order. 
Thus, the magnetoelectric responses in the layered honeycomb lattice are classified according to the applied current direction. 
This relationship is essentially the same as in the insulating case, in which the electric current is replaced by the electric field. 

As shown in Sec.~\ref{sec:ME1}, an electric current in the $z$ direction induces the additional magnitude of the toroidal magnetization to the underlying magnetic order [Eq.~(\ref{eq:M})] with 
preserving the ordering pattern. 
This is a natural consequence of the coupling between toroidal moment and electric current in 
the second term in Eq.~(\ref{eq:H_int}) together with the relation, $\bm{M}_{l}\propto\bm{D}_{l}\times\bm{T}$. 
As suggested by the coupling term, this magnetoelectric effect is present even in the absence of the toroidal order; in fact, the induced magnetization is largest 
for $M_{\bm{T}}=0$ [see the inset of Fig.~\ref{Fig:SAFJZ}(a)]. 

On the contrary, by applying the current in the $x$ ($y$) direction in the toroidal ordered state, the uniform magnetization is induced perpendicular to the current direction within the plane, as shown in Sec.~\ref{sec:ME2}.
This is a transverse response of the magnetization to the electric current, described in the second relation in Eq.~(\ref{eq:PM}). 
As suggested by the relation, this magnetoelectric effect becomes nonzero only in the presence of 
the toroidal magnetization [see the inset of Fig.~\ref{Fig:SAFJZ}(b)]. 
Thus, the results in the metallic state in our model are consistent with the symmetry analysis in Sec.~\ref{Sec:Toroidal moment}. 
The systematic measurements of these magnetoelectric effects will be useful for a detection of toroidal orders in metals.

For comparison, we consider the case of a complementary magnetic order, 
a flux-type one, as shown in the schematic picture in Fig.~\ref{Fig:table_response}. 
This is the magnetic order obtained by rotating the magnetic moments in the toroidal ordered state by $90^\circ$. 
Note that, in the flux state, the magnetic moments are parallel to the antisymmetric vector $\bm{D}_i$ at each site. 
Although both the spatial-inversion and time-reversal symmetries are broken in this flux state as well, the magnetoelectric response within the $xy$ plane appears in a complementary manner to that for the toroidal ordered state, as shown in the table in Fig.~\ref{Fig:table_response}; instead of the transverse response in the toroidal case, 
a longitudinal magnetization is induced by the in-plane current.
These are also consistent with the symmetry analysis in Sec.~\ref{Sec:Toroidal moment}; the longitudinal response is described by the pseudoscalar term in Eq.~(\ref{Eq:ME free energy}). 

\section{Mean-field calculations}
\label{Sec:Mean field calculations}
In Sec.~\ref{Sec:Toroidal ordering in metals}, we simply assumed the toroidal order given by Eq.~(\ref{eq:M}) on the layered honeycomb lattice and discussed the resultant electronic state, transport properties, and magnetoelectric effects. 
Now, we examine when and how such a ferroic toroidal ordered state is realized in the effective single-band model.  
For that purpose, we restore the Coulomb interaction for the mean-field term in Eq.~(\ref{Hubbard_Ham_mean}) in the form  
\begin{align}
{\cal H} =& -t \sum_{\langle i,j \rangle} \sum_{\sigma}
( c_{i \sigma}^{\dagger}c_{j \sigma} + {\rm H.c.}) \nonumber \\
&+ 2\sum_{i} (\bm{s}_{i}\times \bm{D}_i)^{z}
+ U \sum_{i} c_{i \uparrow}^{\dagger} c_{i \uparrow}c_{i \downarrow}^{\dagger}c_{i \downarrow}.
\label{Hubbard_Ham}
\end{align}
Here, we apply the standard Hartree-Fock approximation to the Coulomb $U$ term by assuming the same six-sublattice form of magnetic ordering as $\bm{D}_l$ but allowing arbitrary magnetic 
pattern within the magnetic unit cell (the magnetic moments are assumed to be within the $xy$ plane). 
We calculate the mean fields by taking the sum over $64^3$ grid points in the folded Brillouin zone. 
In Sec.~\ref{Sec:Ground state phase diagram}, we elucidate the ground-state phase diagram. 
Finite-temperature properties are discussed in Sec.~\ref{Sec:Finite temperature property}. 

\subsection{
Ground state}
\label{Sec:Ground state phase diagram}
\begin{figure}[h]
\begin{center}
\includegraphics[width=1.0 \hsize]{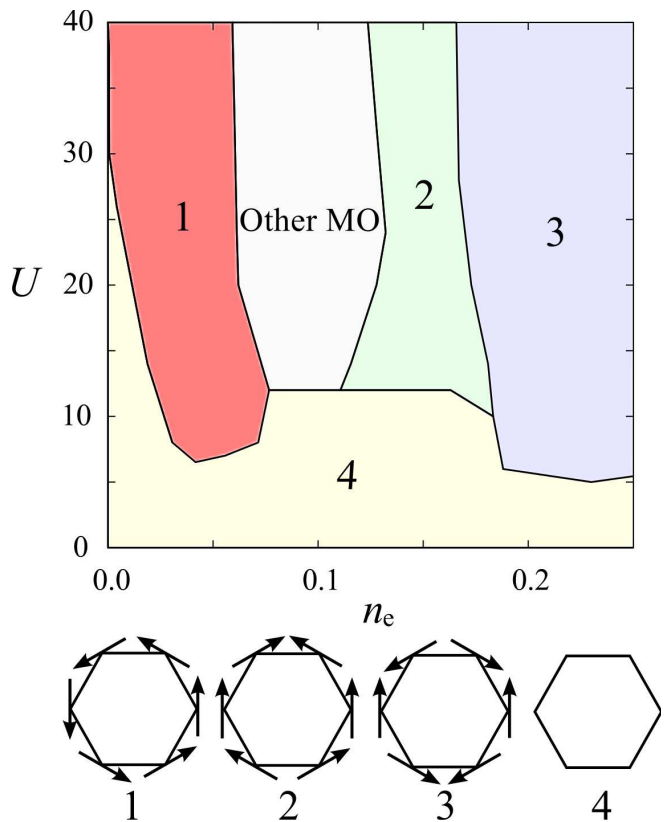}
\caption{
(Color online) 
Ground-state phase diagram of the model in Eq.~(\ref{Hubbard_Ham}) on a layered honeycomb lattice obtained by the mean-field calculations. 
The data are taken at $D=3$. 
Schematic pictures of the ordering patterns are shown in the bottom panel. 
The arrows represent magnetic moments. 
``Other MO'' represents other complicated magnetically ordered states. 
Phase 1 corresponds to the toroidal ordered state. 
There exists no flux-type orders. 
\label{Fig:souzu}
}
\end{center}
\end{figure}

First, we examine the ground state of the model given by Eq.~(\ref{Hubbard_Ham}) by changing $U$ and the electron density $n_{{\rm e}}$. 
Figure~\ref{Fig:souzu} shows the ground-state phase diagram obtained by the mean-field calculations 
at $D=3$.
The result shows that several magnetic states appear in the large $U$ region. 
Among them, the toroidal ordered phase is stabilized in the low-density region. 
This is a metallic state with a shifted band structure and shows the magnetoelectric effects as well as magnetotransport phenomena, as shown in Sec.~\ref{Sec:Toroidal ordering in metals}.

\subsection{
Finite temperature} 
\label{Sec:Finite temperature property}
\begin{figure}[t]
\begin{center}
\includegraphics[width=1.0 \hsize]{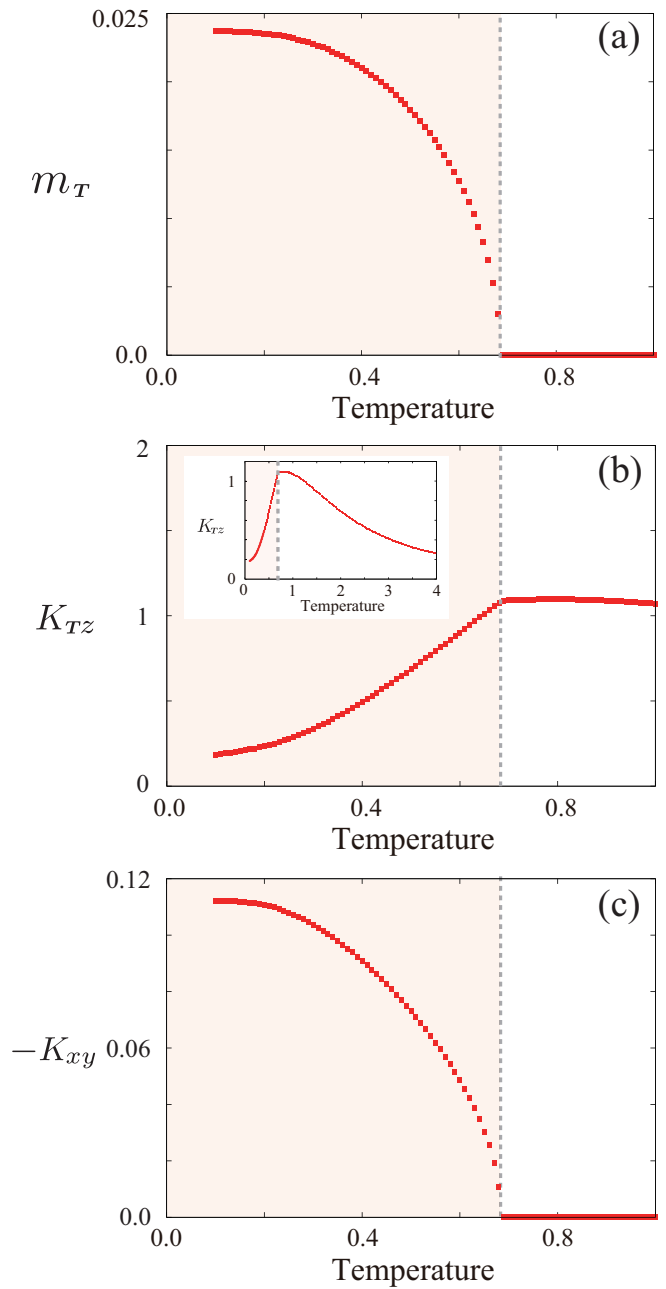} 
\caption{
(Color online) 
Temperature dependence of (a) the magnitude of the toroidal magnetic moments, $m_{\bm{T}}$, 
(b) $K_{\bm{T}z}$, and (c) $K_{xy}$ at $U=40$, $D=3$, and $n_{{\rm e}}=0.05$. 
$K_{\bm{T}z}$ in a broader temperature range is shown in the inset of (b). 
The shaded temperature region indicates the toroidal ordered phase and the vertical dotted lines show the transition temperature $T_c$. 
\label{Fig:Temp_physicalquantity}
}
\end{center}
\end{figure}

Next, we discuss the finite-temperature properties in toroidal ordering within the mean-field approximation. 
The parameters are taken at $U=40$, $D=3$, and $n_{{\rm e}}=0.05$. 
Figure~\ref{Fig:Temp_physicalquantity}(a) shows the result for the magnitude of the toroidal magnetic order parameter, $m_{\bm{T}}=M_{\bm{T}}/(2U)$ 
[see Eq.~(\ref{eq:M}) and Fig.~\ref{Fig:lattice}]. 
While increasing temperature, $m_{\bm{T}}$ decreases continuously to zero at the critical temperature $T_{c}\simeq 0.69$; 
this signals the second-order transition from the low-temperature toroidal ordered state to the high-temperature paramagnetic state. 

Figures~\ref{Fig:Temp_physicalquantity}(b) and \ref{Fig:Temp_physicalquantity}(c) show the temperature dependences of the magnetoelectric responses. 
We use the formula in Eq.~(\ref{Eq:Mangetoelectric effect}) with $\delta=0.1$. 
Figure~\ref{Fig:Temp_physicalquantity}(b) displays $K_{\bm{T}z}$. 
The result shows that the magnitude of $K_{\bm{T}z}$ becomes largest at $T\simeq T_c$ and rapidly decreases for lower temperature after showing a kink at $T_c$ 
[see also the inset in Fig.~\ref{Fig:Temp_physicalquantity}(b)]. 
The behavior is consistent with that expected from the result of the ground state in the inset of Fig.~\ref{Fig:SAFJZ}(a): $K_{\bm{T}z}$ becomes largest at $M_{\bm{T}}=0$ and decreases as $M_{\bm{T}}$ increases. 
From the result, we conclude that the system exhibits a large toroidal magnetic response at and slightly above the critical temperature. 
Note that similar behavior was observed in the magnetoresistance in 
the so-called double-exchange systems, such as 
perovskite manganese oxides~
\cite{Searle1969studies,Tokura1994giant,kaplan1999physics,tokura2000colossal}. 

On the other hand, as shown in Fig.~\ref{Fig:Temp_physicalquantity}(c), 
$-K_{xy}$ behaves like the order parameter $m_{\bm{T}}$; 
it becomes nonzero below $T_c$ and grows rapidly as decreasing temperature. 
This is also consistent with the expectation from the ground-state calculation shown in the inset of Fig.~\ref{Fig:SAFJZ}(b).

\section{Summary and concluding remarks}
\label{Sec:Summary}

In summary, we have investigated the effect and stability of a toroidal order in metals on a lattice without local inversion symmetry. 
We have introduced an effective single-band Hubbard-type model with a site-dependent antisymmetric spin-orbit coupling, while presenting the detailed derivation from a minimal four-band model. 
Considering an in-plane vortex-like magnetic order which accommodates a ferroic toroidal order on a stacked honeycomb lattice, we have studied the effect of the toroidal order on the electronic structure, magnetotransport, and magnetoelectric effects. 
We have explicitly shown in the microscopic model that 
(i) when the toroidal order is realized, the bottom of the electronic bands shifts 
in the direction of the toroidal magnetization, 
(ii) the anisotropic Hall response appears due to the site-dependent antisymmetric spin-orbit coupling, 
(iii) the system exhibits two different types of the magnetoelectric effects: the longitudinal toroidal magnetic response to an electric current in the out-of-plane direction and a transverse uniform magnetization induces by an electric current in the plane. 
We have also investigated the stability of the toroidal ordered state in the effective model by the mean-field approximation. 
We have shown that the toroidal ordered state is stabilized in the strongly-correlated region at low electron density. 
We have also examined the nature of the finite-temperature phase transition for the toroidal ordering 
and the temperature dependence of the magnetoelectric effects. 
We have shown that the transition is continuous and that the toroidal magnetic response is maximized around the critical temperature, while the uniform magnetization induced by current behaves like the toroidal order parameter. 

Our results provide a reference for further exploration of toroidal orders in metallic magnets. 
Our model includes the essential ingredients for toroidal ordering; 
the atomic spin-orbit coupling, off-site hybridizations of different parity orbitals, odd-parity crystalline electric field due to the local inversion symmetry breaking of the lattice structure, and electron-electron correlations. 
A complementary set of measurements of the electronic structure, magnetotransport, and magnetoelectric effects presented here will be useful for identifying the type of toroidal orders. 
We have also mentioned the possibility of magneto-optical effects, such as a nonreciprocal directional dichroism, and of alignment of toroidal domains by cooling the system in an electric current. 

The magnetically ordered state in UNi$_4$B~\cite{Mentink1994, Oyamada2007}
could be a candidate of the spontaneous toroidal ordering discussed in the present study. 
While the lattice structure of this compound is a layered triangular lattice, 
the system shows partial disorder which is the coexistence between the magnetic order on the honeycomb subnetwork and nonmagnetic sites below 20~K. 
Interestingly, the magnetic structure on the stacked honeycomb subnetwork is of vortex-like, as displayed in Fig.~\ref{Fig:lattice}. 
Namely, this compound has the possibility to show the toroidal nature, although the direction and magnitude of the underlying antisymmetric vector $\bm{D}_{l}$ remain unknown. 
Moreover, the electric structure and magnetoelectric effects are not clarified yet, to the best of our knowledge. 
Further experiments, such as the angle-resolved photoemission spectroscopy and the measurement of magnetoelectric tensor, are desirable to examine the possibility of toroidal ordering in this compound. 
Nonlinear optical effect would be interesting as well. 

There are many other candidate materials in which the lattice structure has local inversion symmetry breaking. 
For instance, in spinels, which consist a wide range of compounds including both metals and insulators, the spatial inversion symmetry is broken at the A site; the A sites comprise a diamond lattice, and moreover, each A site locates at the center of a ligand tetrahedron where the local inversion symmetry is lost. 
It is desired to systematically study such materials from the viewpoint of toroidal ordering for further understanding of the exotic electronic and magnetoelectric states.

Finally, let us discuss the possibility of a spontaneous Hall response in the toroidal ordered state. 
We have discussed the magnetotransport and magnetoelectric effects in the toroidal ordered state in Secs.~\ref{Sec:Magnetotransport} and \ref{Sec:Magnetoelectric effect},
respectively. 
By combining these two effects, we deduce that the toroidal order induces an intrinsic Hall response 
even in the absence of an external magnetic field. 
Namely, for example, when we apply the electric field in the $y$ direction, the uniform magnetization is induced in the $x$ direction as shown in Sec.~\ref{Sec:Magnetoelectric effect}, which further induces the electric current in the $z$ direction via the magnetotransport in Sec.~\ref{Sec:Magnetotransport}.  
To clarify such an exotic response, it is necessary to perform the analysis beyond the linear response theory used in the present study. 
Such study is left for future investigation. 

\begin{acknowledgments}
The authors thank T. Arima, H. Harima, J. Nasu, A. Oyamada, and Y. Yanase for fruitful discussions. 
SH is supported by Grant-in-Aid for JSPS Fellows. 
This work was supported by Grants-in-Aid for Scientific Research (No. 24340076), the Strategic Programs for Innovative Research (SPIRE), MEXT, and the Computational Materials Science Initiative (CMSI), Japan. 
\end{acknowledgments}

\appendix
\section{Low-energy effective Hamiltonian 
for the four-band model}
\label{sec:Appendix1}

\begin{figure}[t]
\begin{center}
\includegraphics[width=1.0 \hsize]{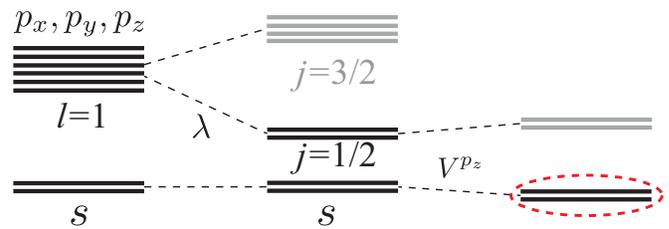} 
\caption{
(Color online) 
(a) Schematic energy diagram in the derivation of the effective single-band model in Eq.~(\ref{Hubbard_Ham_mean}) from the four-band model in Eq.~(\ref{eq:4band_H}). 
The dashed circle represents the energy levels considered in the effective single-band model. 
\label{Fig:effectivemodel}
}
\end{center}
\end{figure}

In this Appendix, we present the derivation of the single-band model in Eq.~(\ref{Hubbard_Ham_mean}). 
It is derived as a low-energy effective model for the four-band model in Eq.~(\ref{eq:4band_H})~\cite{Yanase_kotai}.  
In the derivation, the four-band model is simplified under the following three assumptions. 
First, we consider a 3D system composed of uniform 1D chains in the $z$ direction. 
Namely, we ignore the hopping and hybridization terms between the chains for a while. 
In this 1D limit, from the symmetry, the nonzero matrix element in the hybridization term ${\cal H}_{\rm hyb}$ is only $\tilde{V}_{ij}^{p_z}$ in the $z$ direction. 
We hereafter denote $\tilde{t}_{i,i\pm 1}^\alpha=-t^\alpha$ and $\tilde{V}_{i,i\pm1}^{p_z}=\mp t^{sp_{z}}$ for the nearest neighbors in the $z$ direction. 
Note that the Fourier transform of $\tilde{V}_{i,i\pm1}^{p_z}$ is written in the form 
\begin{align}
V^{p_z} = - 2 {\rm i} \, t^{sp_z} \sin k_z. 
\label{eq:V^p_z_def}
\end{align}
Second, we assume that magnetic moments lie in the $xy$ planes, which are stacked uniformly along the chain. 
Finally, we take the limit of strong spin-orbit coupling $\lambda \rightarrow \infty$. 
In this limit, the sixfold degeneracy in the $p$ orbitals are split into $j=1/2$ doublet and $j=3/2$ quartet, as shown in the middle of Fig.~\ref{Fig:effectivemodel}. 
We take into account 
only the lower $j=1/2$ levels. 
These three assumptions reduce the Hamiltonian for a 1D chain into the form of
\begin{align}
{\mathcal{H}_{{\rm 1D}}}= 
\left( 
\begin{array}{cccc}
\tilde{\epsilon}^s
& \ -\frac{1}{\sqrt{3}}V^{p_z} & \frac{1}{2} \tilde{M}^{-}
& \frac{1}{\sqrt{3}} \tilde{D}^{-}
\\
-\frac{1}{\sqrt{3}}V^{p_z*} & \tilde{\epsilon}^p
-\lambda & -\frac{1}{\sqrt{3}} \tilde{D}^{-}
& \frac{1 }{6} \tilde{M}^{-} \\
\frac{	1}{2} \tilde{M}^{+}
& -\frac{1}{\sqrt{3}} \tilde{D}^{+}
& \tilde{\epsilon}^s & -\frac{1}{\sqrt{3}}V^{p_z}  \\
\frac{1}{\sqrt{3}} \tilde{D}^{+}
&\frac{1}{6} \tilde{M}^{+}
& -\frac{1}{\sqrt{3}}V^{p_z*} & \tilde{\epsilon}^p
- \lambda 
\end{array} \right),  
\label{eq:H_1D_app}
\end{align}
where $\tilde{\epsilon}^s(k_z) = \tilde{E}^s - 2 t^s \cos k_z$, 
$\tilde{\epsilon}^p(k_z) = -\frac{2}{3} (t^{p_x} + t^{p_y} + t^{p_z} ) \cos k_z$, and 
$\tilde{D}^{\pm} = \tilde{D}^{p_x}  \pm {\rm i} \tilde{D}^{p_y}$ and $\tilde{M}^{\pm} = \tilde{M}^{x}  \pm {\rm i} \tilde{M}^{y}$. 
$\tilde{E}^s =\tilde{t}_{ii}^s$ is the atomic energy level for the $s$ orbital. 
Here, we dropped the chain index $i$ for simplicity. 
In Eq.~(\ref{eq:H_1D_app}), we take the basis 
$(\tilde{c}_{
s\uparrow}^{\dagger}, \tilde{c}_{
p +}^{\dagger},\tilde{c}_{
s\downarrow}^{\dagger},\tilde{c}_{
p -}^{\dagger})$, where 
\begin{eqnarray}
\label{eq:p_plus}
\tilde{c}_{
p +}&=& -\frac{1}{\sqrt{3}} \left( \tilde{c}_{
p_x \downarrow}+{\rm i} \tilde{c}_{
p_y \downarrow} + \tilde{c}_{
p_z \uparrow} \right), \\
\label{eq:p_minus}
\tilde{c}_{
p -}&=& \frac{1}{\sqrt{3}} \left( \tilde{c}_{
p_x \uparrow}-{\rm i} \tilde{c}_{ 
p_y \uparrow} - \tilde{c}_{ 
p_z \downarrow} \right). 
\end{eqnarray}

From Eq.~(\ref{eq:H_1D_app}), we project out the lowest two levels after taking into account the hybridization between $s$ and $j=1/2$ levels, $V^{p_z}$. 
Namely, we rewrite the Hamiltonian matrix in Eq.~(\ref{eq:H_1D_app}) in terms of the basis which diagonalizes the upper-left and lower-right 2$\times$2 matrices in Eq.~(\ref{eq:H_1D_app}). 
In the new basis, we take into account only the two lower-energy levels; 
the off-diagonal components between these two levels give the effective antisymmetric spin-orbit coupling and the electron-electron interaction in a mean-field form. 
Thus, we end up with an effective single-band model, whose Hamiltonian is given by 
\begin{align}
\label{eq:H_1D_x}
\mathcal{H}_{\rm 1D} 
&=  \sum_{\sigma} 
\epsilon(k_z) 
c_{\sigma}^{\dagger} c_{\sigma} +
2 (\bm{s}\times \bm{D}(k_z))^{z} - \bm{M}(k_z)\cdot\bm{s}, 
\end{align}
where ($c_{\sigma}$, $c_{\sigma}^\dagger$) is a new basis after considering the effect of the hybridization, $V^{p_z}$, and $\bm{s} = \sum_{\sigma,\sigma'} c_\sigma^\dagger (\bm{\sigma}_{\sigma\sigma'}/2) c_\sigma'$.  
$\sigma$ is the quasispin distinguishing the two low-energy time-reversal pair states. 
Here, 
\begin{align}
\label{eq:E}
\epsilon (k_z)
&= 
\tilde{\epsilon}_+(k_z) 
- E(k_z)
=\epsilon(-k_{z}),\\
\label{eq:DVp}
\bm{D}
(k_z) &= \frac{1}{\sqrt{3}} \frac{\tilde{t}^{sp_z}  \sin k_z}{ 
E(k_z)} \tilde{\bm{D}}
=-\bm{D}(-k_{z}), 
\\
\label{eq:MVp}
\bm{M}(k_z)
&=\frac{2}{3} \left[ 1- \frac{1}{2} 
\frac{\tilde{\epsilon}_-(k_z)
}{E(k_z)} \right] \tilde{\bm{M}}
=\bm{M}(-k_{z}), 
\end{align}
and 
\begin{align}
E(k_z) &= [
(\tilde{\epsilon}_-(k_z))^2 
+(\tilde{t}^{sp_z} \sin k_z)^2]^{1/2},
\\
\tilde{\epsilon}_{\pm}(k_z) &= \frac12 
[\tilde{\epsilon}^s (k_z) 
\pm ( \tilde{\epsilon}^p (k_z) -\lambda) ], 
\\
\tilde{t}^{sp_z} &= \frac{2}{\sqrt{3}}t^{sp_z}.
\end{align} 
We assume that $\tilde{\bm{D}} = (\tilde{D}^{p_x},\tilde{D}^{p_y},0)$ and $\tilde{\bm{M}} = (\tilde{M}^x,\tilde{M}^y,0)$, which lie in the $xy$ plane and are parallel to $\bm{D}$ and $\bm{M}$, respectively. 
It should be noted that the antisymmetric vector $\bm{D}(k_{z})$ in Eq.~(\ref{eq:DVp}) involves the off-site hybridization $t^{sp_z}$ and the odd-parity crystalline electric field $\tilde{\bm{D}}$ under the assumption of the strong atomic spin-orbit coupling. 

Finally, we simplify the $k_z$ dependence in Eq.~(\ref{eq:H_1D_x}) as follows. 
For $\bm{D}(k_z)$, hereafter we retain the most interesting part, $\sin k_z$ in the numerator, for simplicity. 
We also drop all the $k_z$ dependence of $\bm{M}(k_z)$. 
Examples of the simplified forms of $\bm{D}$ and $\bm{M}$ are found in Eqs.~(\ref{eq:D}) and (\ref{eq:M}). 
For the dispersion in Eq.~(\ref{eq:E}), we replace it by a simple form $\epsilon(k_z) = -2t \cos k_z$ with a renormalized transfer integral $t$. 
These simplifications on Eq.~(\ref{eq:H_1D_x}) lead to the effective single-band Hamiltonian for a single chain, $\mathcal{H}_{{\rm 1D}}^{{\rm MF}}$ in Eq.~(\ref{eq:localeff}). 
Finally, we obtain the effective single-band Hamiltonian $\mathcal{H}^{\rm MF} 
$ in Eq.~(\ref{Hubbard_Ham_mean}) by restoring the hopping term in the $xy$ plane neglected in the derivation.


\end{document}